\documentclass[pra,twocolumn]{revtex4}
\usepackage{amssymb}
\usepackage{amsfonts}
\usepackage{amsmath}
\usepackage{graphicx}

\begin{document}

\title{Conditional past-future correlation in dephasing environments}
\title{Break of conditional past-future independence induced by
non-Markovian dephasing baths}
\title{Conditional past-future correlation induced by non-Markovian
dephasing reservoirs}
\author{Adri\'{a}n A. Budini}
\affiliation{Consejo Nacional de Investigaciones Cient\'{\i}ficas y T\'{e}cnicas
(CONICET), Centro At\'{o}mico Bariloche, Avenida E. Bustillo Km 9.5, (8400)
Bariloche, Argentina, and Universidad Tecnol\'{o}gica Nacional (UTN-FRBA),
Fanny Newbery 111, (8400) Bariloche, Argentina}
\date{\today}

\begin{abstract}
Memory effects can be studied through a conditional past-future correlation,
which measures departure with respect to a conditional past-future
independence valid in a memoryless Markovian regime. In a quantum regime
this property leads to an operational definition of quantum non-Markovianity
based on three consecutive system measurement processes and postselection
[Budini, Phys. Rev. Lett. \textbf{121}, 240401 (2018)]. Here, we study the
conditional past-future correlation for a qubit system coupled to different
dephasing environments. Exact solutions are obtained for a quantum spin bath
as well as for classically fluctuating random Hamiltonian models. The
developing of memory effects and departures from Born-Markov or white-noise
approximations are related to a measurement back action that changes the
system dynamics between consecutive measurements. It is shown that this
effect may develop even when the former system evolution is given by a
time-independent Lindblad equation. This unusual non-Markovian case arises
when the characteristic parameters of the dynamics become Lorentzian random
distributed variables. 
\end{abstract}

\maketitle

\section{Introduction}

In a classical regime, Markovianity (memoryless property) leads to
descriptions based on local-in-time evolutions such as Fokker-Planck
equations and master equations \cite{vanKampen}. In a quantum regime,
instead of probabilities, a density matrix operator describes the open
system dynamics \cite{breuerbook,vega}. As is well known, when a
local-in-time description applies the evolution of the density matrix must
to assume a Lindblad structure \cite{alicki}. Therefore, in the last years
these equations were naturally related to quantum Markovianity. In fact,
different quantum memory measures (quantum non-Markovianity measures)\ rely
on diverse departures that a system may develops with respect to their
properties \cite{BreuerReview,plenioReview}. Many alternative proposals were
studied \cite%
{BreuerFirst,cirac,rivas,DarioSabrina,BaeDario,dario11,Acin,cresser,canonicalCresser,geometrical,fisher,mutual,brasil,fidelity,eternal,maximal}%
, most of them based on the behavior of different quantum information
measures under a Lindblad evolution.

Usually in the definition of the previous memory indicators the only
available information is given by the density matrix evolution or
propagator. Memory effects developing in open quantum systems can also be
defined on alternative grounds. For example, the well-established notion of
classical\ Markovianity \cite{vanKampen} can be extended to a quantum regime
by subjecting the system to extra control operations (measurements) \cite%
{modi,PRL}. The operational definition of quantum Markovianity introduced in
Ref.~\cite{modi} is based on a \textquotedblleft process
tensor\textquotedblright\ framework, which relies on the usual definition of
classical Markovianity in terms of conditional probability distributions.
Thus, quantum Markovianity is defined by a conditional independence of the
system dynamics on past control operations.

The formalism of Ref. \cite{PRL} relies on an equivalent but different
formulation of classical Markovianity, that is, \textit{the\ statistical
independence of past and future events when conditioned to a given state at
the present time}~\cite{CoverTomas}. Equivalently, memory effects break
conditional past-future (CPF) independence. Hence, an ensemble of three
time-ordered (random) system events provides a minimal basis for
establishing classical and quantum Markovianity. A related conditional
past-future correlation becomes an univocal indicator of departures from a
memoryless regime. In a quantum regime, the three events correspond to the
outcomes of three (system) measurement processes. \textit{Postselection}
take into account the conditional character of the definition. The CPF\
correlation vanishes whenever a Born-Markov or white noise approximation
applies to quantum or classical environments respectively \cite{PRL}. Its
calculation involves both predictive and retrodicted quantum probabilities 
\cite{vaidman,murch,molmer}. Hence, techniques and concepts coming from
retrodicted quantum measurements \cite%
{vaidman,murch,molmer,haroche,huard,naghi,decay,retro,wiseman,dressel,barnett}
play a fundamental role in this alternative approach.

In this paper we study the CPF correlation for a qubit system interacting
with different non-Markovian dephasing environments such as a quantum spin
bath and stochastic Hamiltonian models. An explicit derivation complement
the exact results presented in Ref.~\cite{PRL}. In addition, the developing
of memory effects and departure from Born-Markov and white noise
approximations are studied in detail for both kinds of models. These
features are related to a measurement back action that \textit{changes the
system dynamics between consecutive measurements processes. }Contrarily to
all previous non-Markovian measures \cite{BreuerReview,plenioReview}, we
explicitly show that even\ when a time-independent Lindblad equation defines
the former system evolution (between the first and second measurements), its
posterior dynamics (between the second and third measurements) can be
modified. This unusual non-Markovian dynamics arises in both kinds of models
when the underlying parameters become random (time-independent) variables
characterized by a Lorentzian probability density.

The paper is outlined as follows. In Sec. II we briefly resume the formalism
established in Ref. \cite{PRL}. In Sec.~III the spin bath model is studied.
Sec. IV is devoted to stochastic Hamiltonian dynamics. In Sec. V, both kinds
of models are characterized when the underlying parameters become Lorentzian
random variables. The conclusions are provided in Sec. VI.

\section{Conditional past-future correlation}

Given an ensemble of three time ordered random events $x\rightarrow
y\rightarrow z$ occurring at times $t_{x}<t_{y}<t_{z},$ Bayes rule allow us
to write the probability $P(z,x|y)$ of past $(x)$ and future events $(z),$
conditioned to a given present state $(y),$ as%
\begin{equation}
P(z,x|y)=P(z|y,x)P(x|y),  \label{Indenpendence}
\end{equation}%
where in general $P(b|a)$ denotes the conditional probability of $b$ given $%
a.$ A \textit{conditional past-future correlation},%
\begin{equation}
C_{pf}=\langle O_{z}O_{x}\rangle _{y}-\langle O_{z}\rangle _{y}\langle
O_{x}\rangle _{y},
\end{equation}%
defined as%
\begin{equation}
C_{pf}=\sum_{zx}[P(z,x|y)-P(z|y)P(x|y)]O_{z}O_{x},  \label{CPFExplicit}
\end{equation}%
is a measure of memory non-Markovian effects \cite{PRL}. In fact, for
Markovian processes $P(z|y,x)=P(z|y),$ implying $P(z,x|y)=P(z|y)P(x|y).$
Thus, $C_{pf}=0.$ In Eq.~(\ref{CPFExplicit}) the sum indexes $z$ and $x$ run
over all possible outcomes occurring at times $t_{z}$ and $t_{x}$
respectively, while $y$ is a fixed particular possible value at time $t_{y}.$
The parameters $\{O_{z}\}$ and $\{O_{x}\}$ correspond to a property
associated to each system state.

In a quantum regime, the sequence $x\rightarrow y\rightarrow z$ is given by
the outcomes of three consecutive measurements performed over the system of
interest. The corresponding measurement operators \cite{milburn} are $%
x\rightarrow \Omega _{x},$ $y\rightarrow \Omega _{y},$ $z\rightarrow \Omega
_{z},$ and fulfill $\sum_{x}\Omega _{x}^{\dag }\Omega _{x}=\sum_{y}\Omega
_{y}^{\dag }\Omega _{y}=\sum_{z}\Omega _{z}^{\dag }\Omega _{z}=\mathrm{I},$
where $\mathrm{I}$ is the identity matrix in the system Hilbert space and
the sum indexes run over all possible outcomes at each stage. Furthermore,
in Eq. (\ref{CPFExplicit}) $\{O_{z}\}$ and $\{O_{x}\}$ are set by the
measured observables. Given the conditional character of the past-future
correlation, in an experimental setup it follows from a postselected
sub-ensemble of realizations where the intermediate $y$-outcome is a fixed
arbitrary one. On the other hand, the calculation of $P(z|y,x)$ relies on
standard predictive quantum measurement theory. In contrast, $P(x|y)$ is a
retrodicted probability that can be read from a \textquotedblleft past
quantum state\textquotedblright\ formalism \cite{molmer,retro}.

\section{Dephasing spin bath}

Similarly to Ref. \cite{PRL}, here we consider a qubit system interacting
with\ a quantum spin bath \cite{Zurek,ZurekRMP,Paz}. Their mutual
interaction is set by the Hamiltonian%
\begin{equation}
H_{T}=\sigma _{\hat{z}}\otimes \sum_{k=1}^{N}g_{k}\sigma _{\hat{z}}^{(k)}.
\label{SpinBath}
\end{equation}%
In here, $\sigma _{\hat{z}}$ is the system Pauli matrix in the $\hat{z}-$%
direction. Its eigenvectors are denoted as $|\pm \rangle .$ On the other
hand, $\sigma _{\hat{z}}^{(k)}$ is the $\hat{z}-$Pauli matrix corresponding
to the $k$-spin, whose eigenvectors are denoted as $|\uparrow \rangle _{k}$
and $|\downarrow \rangle _{k}.$ $\{g_{k}\}$ measure the coupling between
each spin of the environment and the qubit.

The interaction model (\ref{SpinBath}) always admits an exact solution \cite%
{Zurek,ZurekRMP,Paz}. For a separable pure initial bipartite state, $\rho
_{0}^{se}=|\Psi _{0}\rangle \langle \Psi _{0}|,$ with%
\begin{equation}
|\Psi _{0}\rangle =(a|+\rangle +b|-\rangle )\otimes \sum_{k=1}^{N}(\alpha
_{k}|\uparrow \rangle _{k}+\beta _{k}|\downarrow \rangle _{k}),
\label{PsiInitial}
\end{equation}%
where the initial bath state is sets by the individual spin coefficients $%
\{\alpha _{k}\}$ and $\{\beta _{k}\},$ at time $t$ the bipartite state is $%
\rho _{t}^{se}=|\Psi _{t}\rangle \langle \Psi _{t}|,$ where%
\begin{equation}
|\Psi _{t}\rangle =a|+\rangle \otimes |\mathcal{B}(t)\rangle +b|-\rangle
\otimes |\mathcal{B}(-t)\rangle .  \label{Psi(t)}
\end{equation}%
Thus, the system and the environment become entangled. The normalized bath
state $[\langle \mathcal{B}(t)|\mathcal{B}(t)\rangle =1]$ is \cite%
{ZurekRMP,Paz}%
\begin{equation}
|\mathcal{B}(t)\rangle =\prod_{k=1}^{N}(\alpha _{k}e^{+i2g_{k}t}|\uparrow
\rangle +\beta _{k}e^{-i2g_{k}t}|\downarrow \rangle ).  \label{B(t)}
\end{equation}

The three measurement processes that define the CPF correlation [Eq.~(\ref%
{CPFExplicit})] are chosen as projective ones, all of them being performed
in $\hat{x}$-direction of the qubit Bloch sphere. Thus, the outcomes of each
measurement, in successive order, are $x=\pm 1,$ $y=\pm 1,$ and $z=\pm 1,$
which in turn define the system operators values $O_{z}=z$ and $O_{x}=x.$
The corresponding measurement operators are the same, $\{\Omega
_{x}\}=\{\Omega _{y}\}=\{\Omega _{z}\}=\{\Pi _{\hat{x}=\pm 1}\},$ where $\Pi
_{\hat{x}=\pm 1}=|\hat{x}_{\pm }\rangle \langle \hat{x}_{\pm }|,$ with $|%
\hat{x}_{\pm }\rangle =(|+\rangle \pm |-\rangle )/\sqrt{2}.$ A hat symbol
distinguishes directions in Bloch sphere from measurement outcomes.

\subsection{Conditional probabilities}

In the following calculations the initial system state is $|+\rangle $ [Eq.~(%
\ref{PsiInitial}) with $a=1$ and $b=0].$ Thus,%
\begin{equation}
|\Psi _{0}\rangle =|+\rangle \otimes \sum_{k=1}^{N}(\alpha _{k}|\uparrow
\rangle _{k}+\beta _{k}|\downarrow \rangle _{k}),  \label{InitialPosta}
\end{equation}%
is the initial bipartite\ system-environment state. The goal is to calculate 
$P(x|y)$ and $P(z|y,x)$ [Eq. (\ref{Indenpendence})].

At all steps the bipartite state remains a pure one. After the first $x$%
-measurement, from standard quantum measurement theory \cite{milburn}, it
becomes $|\Psi _{0}\rangle \rightarrow |\Psi _{0}^{x}\rangle =\Pi _{\hat{x}%
=x}|\Psi _{0}\rangle /\sqrt{\langle \Psi _{0}|\Pi _{\hat{x}=x}|\Psi
_{0}\rangle },$ delivering%
\begin{equation}
|\Psi _{0}^{x}\rangle =\frac{|+\rangle +x|-\rangle }{\sqrt{2}}\otimes
\sum_{k=1}^{N}(\alpha _{k}|\uparrow \rangle _{k}+\beta _{k}|\downarrow
\rangle _{k}),  \label{Psi0X}
\end{equation}%
where consistently $x=\pm 1$ is the outcome of the first measurement. The
probability of both options is $P(x)=\langle \Psi _{0}|\Pi _{\hat{x}=x}|\Psi
_{0}\rangle =1/2.$

After evolving up to a time $t\equiv t_{y}-t_{x},$ from Eq. (\ref{Psi(t)})
the bipartite state becomes%
\begin{equation}
|\Psi _{t}^{x}\rangle =\frac{1}{\sqrt{2}}(|+\rangle \otimes |\mathcal{B}%
(t)\rangle +x|-\rangle \otimes |\mathcal{B}(-t)\rangle ).  \label{PsiTiempo}
\end{equation}%
Posteriorly, the second measurement, with outcomes $y=\pm 1,$ is performed
in $\hat{x}-$direction.\ The probability of each option, given that the
previous outcome was $x,$ is $P(y|x)=\langle \Psi _{t}^{x}|\Pi _{\hat{x}%
=y}|\Psi _{t}^{x}\rangle ,$ which delivers $P(y|x)=(1+yx\mathrm{Re}[\langle 
\mathcal{B}(-t)|\mathcal{B}(t)\rangle ])/2.$ Introducing the joint
probability of both outcomes $P(y,x)=P(y|x)P(x),$ 
\begin{equation}
P(y,x)=\frac{1}{4}(1+yx\mathrm{Re}[\langle \mathcal{B}(-t)|\mathcal{B}%
(t)\rangle ]),  \label{ConjuntaYXSpin}
\end{equation}%
it follows $P(y)=\sum_{x=\pm 1}P(y,x)=1/2.$ The retrodicted probability $%
P(x|y)=P(y,x)/P(y)$ then reads%
\begin{equation}
P(x|y)=\frac{1}{2}(1+yx\mathrm{Re}[\langle \mathcal{B}(-t)|\mathcal{B}%
(t)\rangle ]).  \label{Pxy}
\end{equation}%
Due to the chosen system initial condition, it follows the symmetry $%
P(x|y)=P(y|x).$

After the second $y$-measurement the bipartite state change as $|\Psi
_{t}^{x}\rangle \rightarrow |\Psi _{t}^{yx}\rangle =\Pi _{\hat{x}=y}|\Psi
_{t}^{x}\rangle /\sqrt{\langle \Psi _{t}^{x}|\Pi _{\hat{x}=y}|\Psi
_{t}^{x}\rangle },$ which from Eq. (\ref{PsiTiempo}) reads%
\begin{equation}
|\Psi _{t}^{yx}\rangle =\frac{|+\rangle +y|-\rangle }{\sqrt{2}}\otimes |%
\mathcal{B}_{yx}(t)\rangle .  \label{Uncorrelato}
\end{equation}%
The bath state is%
\begin{equation}
|\mathcal{B}_{yx}(t)\rangle \equiv \frac{|\mathcal{B}(t)\rangle +yx|\mathcal{%
B}(-t)\rangle }{\sqrt{\mathcal{N}_{t}^{yx}}},  \label{Byx}
\end{equation}%
with normalization constant $\mathcal{N}_{t}^{yx}=\langle \mathcal{B}(t)|%
\mathcal{B}(t)\rangle +\langle \mathcal{B}(-t)|\mathcal{B}(-t)\rangle
+yx(\langle \mathcal{B}(t)|\mathcal{B}(-t)\rangle +\langle \mathcal{B}(-t)|%
\mathcal{B}(t)\rangle ).$ It can be rewritten as $\mathcal{N}%
_{t}^{yx}=4P(x|y)$ [Eq. (\ref{Pxy})].

In the next step, the bipartite arrangement evolves during a time interval $%
\tau \equiv t_{z}-t_{y}$ with the unitary dynamics dictated by the
Hamiltonian (\ref{SpinBath}), $|\Psi _{t}^{yx}\rangle \rightarrow |\Psi
_{t+\tau }^{yx}\rangle .$ From Eq.~(\ref{Psi(t)}) it follows%
\begin{equation}
|\Psi _{t+\tau }^{yx}\rangle =\frac{1}{\sqrt{2}}(|+\rangle \otimes |\mathcal{%
B}_{yx}(t+\tau )\rangle +y|-\rangle \otimes |\mathcal{B}_{yx}(t-\tau
)\rangle ).  \label{PsiXY}
\end{equation}%
The probability of the last $z$-measurement $P(z|y,x)=\langle \Psi _{t+\tau
}^{yx}|\Pi _{\hat{x}=z}|\Psi _{t+\tau }^{yx}\rangle ,$ reads%
\begin{equation}
P(z|y,x)=\frac{1}{2}(1+zy\mathrm{Re}[\langle \mathcal{B}_{yx}(t-\tau )|%
\mathcal{B}_{yx}(t+\tau )\rangle ]).  \label{Pz|yx}
\end{equation}%
Eqs. (\ref{Pxy}) and (\ref{Pz|yx}) are the central results of this
subsection.

\subsection{Dynamics between consecutive measurements}

The previous analysis allows us to characterize the system dynamics between
consecutive measurement events. After the first $x$-measurement and before
the second $y$-measurement [time interval $(0,t)],$ the system state follows
as $\rho _{t}^{x}=\mathrm{Tr}_{e}[|\Psi _{t}^{x}\rangle \langle \Psi
_{t}^{x}|],$ where $|\Psi _{t}^{x}\rangle $ is given by Eq.~(\ref{PsiTiempo}%
), and $\mathrm{Tr}_{e}[\cdot ]$ is the trace operation over the environment
degrees of freedom. We get,%
\begin{equation}
\rho _{t}^{x}=\frac{1}{2}\left( 
\begin{array}{cc}
1 & xc_{t} \\ 
xc_{t}^{\ast } & 1%
\end{array}%
\right) ,  \label{RhoX}
\end{equation}%
where the coherence behavior, from Eq. (\ref{B(t)}), is given by%
\begin{equation}
c_{t}\equiv \langle \mathcal{B}(-t)|\mathcal{B}(t)\rangle
=\prod_{k=1}^{N}(|\alpha _{k}|^{2}e^{+i2g_{k}t}+|\beta
_{k}|^{2}e^{-i2g_{k}t}).  \label{CoherTimeMicro}
\end{equation}%
Consistently with the underlying interaction [Eq. (\ref{SpinBath})], only
the system coherences are affected.

After the second $y$-measurement and before the third $z$-measurement [time
interval $(t,t+\tau )],$ the system state follows as $\rho _{t,\tau }^{yx}=%
\mathrm{Tr}_{e}[|\Psi _{t+\tau }^{yx}\rangle \langle \Psi _{t+\tau }^{yx}|],$
where $|\Psi _{t+\tau }^{yx}\rangle $ is given by Eq.~(\ref{PsiXY}). We get,%
\begin{equation}
\rho _{t,\tau }^{yx}=\frac{1}{2}\left( 
\begin{array}{cc}
1 & yc_{t,\tau }^{yx} \\ 
yc_{t,\tau }^{yx\ast } & 1%
\end{array}%
\right) ,
\end{equation}%
where the new coherence behavior $c_{t,\tau }^{yx},$ from Eq. (\ref{Byx}),
is given by%
\begin{equation}
c_{t,\tau }^{yx}\equiv \langle \mathcal{B}_{yx}(t-\tau )|\mathcal{B}%
_{yx}(t+\tau )\rangle \!=\!\frac{c_{\tau }+yx(c_{t+\tau }+c_{t-\tau }^{\ast
})/2}{1+yx(c_{t}+c_{t}^{\ast })/2}.  \label{CtTau}
\end{equation}%
Here, $c_{t}$ gives the previous coherence behavior, Eq. (\ref%
{CoherTimeMicro}). In contrast, $c_{t,\tau }^{yx}$ explicitly depends on
both previous measurement results.

From Eqs. (\ref{PsiTiempo}) and (\ref{PsiXY}), it is evident that for this
model a Born-Markov approximation \cite{breuerbook,vega} does not applies at
any stage (separable system-bath state). \textit{This non-Markovian property
can also be read from a measurement back action that leads to a change of
system dynamics between consecutive measurements,} $c_{t,\tau }^{yx}\neq
c_{t}.$

The former bipartite dynamics in $(0,t)$ begins in a separable state [Eq. (%
\ref{Psi0X})]. Due to the projective nature of the second $y$-measurement,
this property is also valid for the interval $(t,t+\tau )$ [Eq. (\ref%
{Uncorrelato})]. Nevertheless, in contrast here the bath state $|\mathcal{B}%
_{yx}(t)\rangle $ is an entangled one that involves all spin bath variables
[Eq. (\ref{Byx})]. It is a superposition of the bath states $|\mathcal{B}%
(t)\rangle $ and $|\mathcal{B}(-t)\rangle $ whose phase in turn depends on
the product of outcomes $yx.$ This measurement back action on the bath
degrees of freedom leads to a different posterior system dynamics. Thus,
this change can in fact be read as a fingerprint of non-Markovian effects
and departure from Born-Markov approximation.

\subsection{CPF\ correlation}

The CPF correlation (\ref{CPFExplicit}) can be calculated after getting the
CPF probability $P(z,x|y)=P(z|y,x)P(x|y).$ From Eqs. (\ref{Pxy}) and (\ref%
{Pz|yx}), jointly with Eqs. (\ref{CoherTimeMicro}) and (\ref{CtTau}), it
follows%
\begin{eqnarray}
P(z,x|y) &=&\frac{1}{4}\Big{\{}1+xyf(t)+zyf(\tau )  \notag \\
&&+zx\frac{[f(t+\tau )+f(t-\tau )]}{2}\Big{\}},  \label{CPFProbability}
\end{eqnarray}%
where for simplifying the expression we defined $f(t)\equiv \mathrm{Re}%
[c_{t}].$ In addition, 
\begin{subequations}
\begin{eqnarray}
P(z|y) &=&\sum_{x=\pm 1}P(z,x|y)=\frac{1}{2}[1+zyf(\tau )], \\
P(x|y) &=&\sum_{z=\pm 1}P(z,x|y)=\frac{1}{2}[1+xyf(t)].
\end{eqnarray}%
The conditional averages then reads $\langle O_{z}\rangle _{y}=yf(\tau ),$ $%
\langle O_{x}\rangle _{y}=yf(t),$ while from Eq.~(\ref{CPFProbability}) we
get $\langle O_{z}O_{x}\rangle _{y}=[f(t+\tau )+f(t-\tau )]/2,$ where it has
been used that $O_{z}=z,$ $O_{x}=x.$ The exact expression for the CPF
correlation (\ref{CPFExplicit}) $[C_{pf}\rightarrow C_{pf}(t,\tau )]$ then
is 
\end{subequations}
\begin{equation}
C_{pf}(t,\tau )=\frac{f(t+\tau )+f(t-\tau )}{2}-f(t)f(\tau ).
\label{CPFSpin}
\end{equation}%
This result recovers the exact expression presented in Ref.~\cite{PRL}. 
\begin{figure}[tbp]
\includegraphics[bb=45 875 735 1132,angle=0,width=8.8cm]{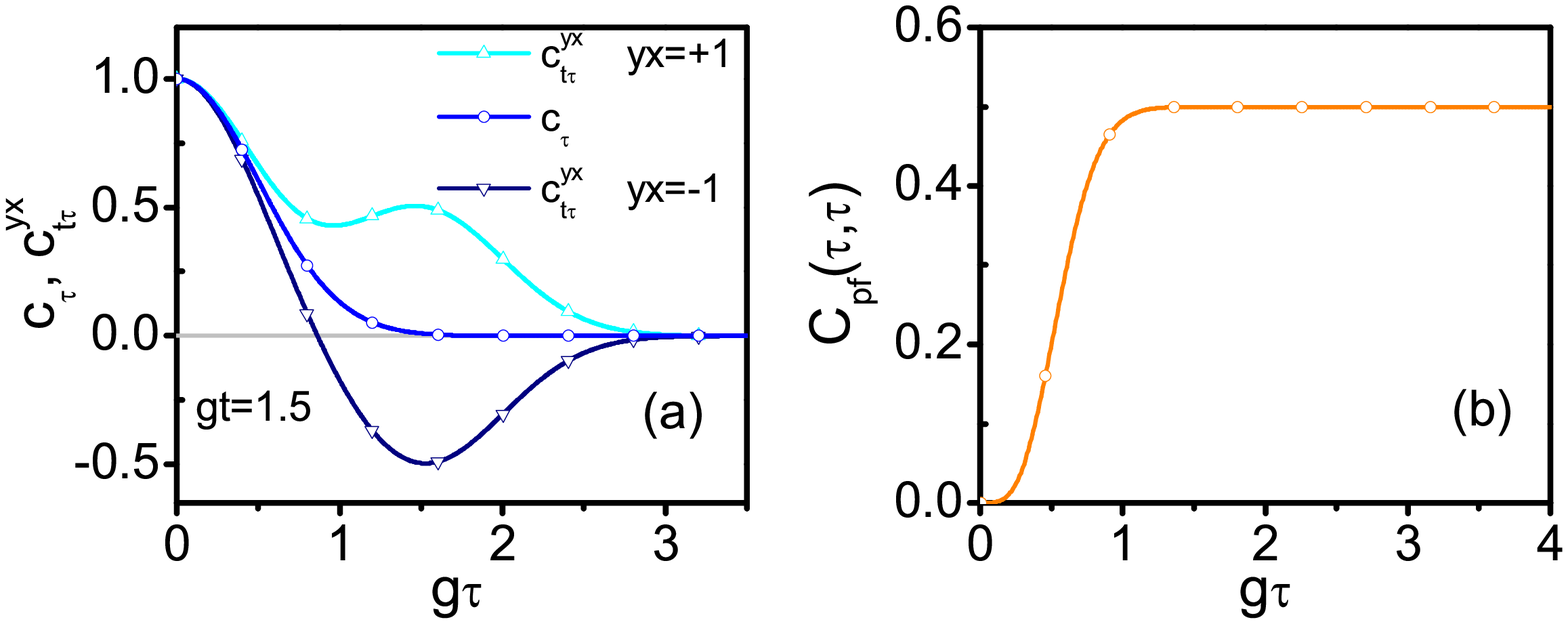}
\caption{(a) System coherence between consecutive measurements $c_{t}$ [Eq. (%
\protect\ref{CoherTimeMicro})] and $c_{t,\protect\tau }^{yx}$\ [Eq. (\protect
\ref{CtTau})] for the spin bath model. (b) CPF correlation (\protect\ref%
{CPFSpin}) for equal times, $C_{pf}(\protect\tau ,\protect\tau ).$ In both
cases, $N=50,$ the coupling are given by the scaling Eq. (\protect\ref%
{Scaling}), while $\protect\alpha _{k}=\protect\beta _{k}=1/2.$ The system
begins in the state $|+\rangle .$}
\end{figure}

\subsection{Example}

In order to exemplify the previous analysis, we consider a regime where the
spin bath model leads to Gaussian system decay behaviors \cite{Paz},
situation that in turn is of interest in different experimental situations 
\cite{pasta}.

All spins starts in the same state, $\alpha _{k}=\alpha ,$ $\beta _{k}=\beta
,$ with $|\alpha |^{2}+|\beta |^{2}=1,$ and $g_{k}=g_{N}.$ Thus, the system
coherence behavior after the first $x$-measurement [Eq. (\ref{CoherTimeMicro}%
)]\ becomes%
\begin{equation}
c_{t}=(|\alpha |^{2}e^{+i2g_{N}t}+|\beta |^{2}e^{-i2g_{N}t})^{N}.
\label{ctExplicita}
\end{equation}%
In order to obtain an asymptotic behavior $(N\gg 1)$ independent of $N$, the
following scaling is assumed%
\begin{equation}
g_{N}=\frac{1}{\sqrt{N}}g,\ \ \ \ \ \ \ \ \ \ |\alpha |^{2}-|\beta |^{2}=%
\frac{\omega }{2g\sqrt{N}},  \label{Scaling}
\end{equation}%
where $g$ and $\omega $ are free parameters. In the limit $N\gg 1,$ from Eq.
(\ref{ctExplicita}) we get%
\begin{equation}
c_{t}\simeq \exp [+i\omega t-2(gt)^{2}].  \label{Gauss1}
\end{equation}%
On the other hand, the coherence behavior after the second $y$-measurement,
given by $c_{t,\tau }^{yx}$ [Eq. (\ref{CtTau})], can be straightforwardly
approximated from this expression. When $|\alpha |^{2}=|\beta |^{2}=1/2,$ $%
c_{t}$\ follows a pure Gaussian decay behavior $[\omega =0]$ while%
\begin{equation}
c_{t,\tau }^{yx}\simeq \frac{e^{-2(g\tau )^{2}}+yx[e^{-2g^{2}(t+\tau
)^{2}}+e^{-2g^{2}(t-\tau )^{2}}]/2}{1+yxe^{-2(gt)^{2}}}.  \label{Gauss2}
\end{equation}
\begin{figure}[tbp]
\includegraphics[bb=0 0 330 250,angle=0,width=6cm]{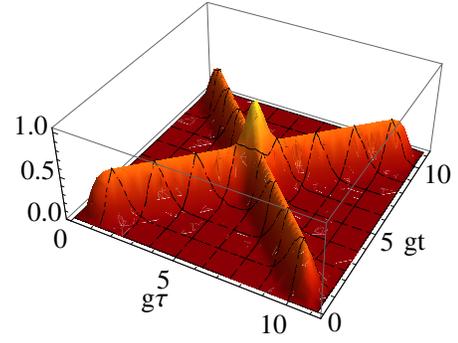}
\caption{CPF correlation $C_{pf}(t,\protect\tau )$ [Eq. (\protect\ref%
{CPFSpin})] as a function of both measurements times. The parameters are the
same than in the previous figure. The behavior is periodic in both
measurement time-intervals, $t$ and $\protect\tau $. For increasing number
of bath spins, $N\rightarrow \infty ,$ the location of the central peak
diverges.}
\end{figure}

Taking the scaling defined in Eq. (\ref{Scaling}), in Fig. 1 we plot the
system coherence behaviors between consecutive measurements, $c_{t}$ [Eq. (%
\ref{CoherTimeMicro})] and $c_{t,\tau }^{yx}$ [Eq. (\ref{CtTau})]. Both
objects are very well fitted by Eqs. (\ref{Gauss1}) and (\ref{Gauss2})
respectively. On the other hand, we note that in general $c_{t,\tau }^{yx},$
as a function of $\tau ,$ may develops strong departures with respect to $%
c_{t}.$ This feature depends on the product of outcomes $yx$ and is induced
by the measurements back action that lead to different \textquotedblleft
initial\textquotedblright\ bath states, Eqs.~(\ref{Psi0X}) and (\ref%
{Uncorrelato}) respectively. This property in turn lead to strong
non-Markovian effects, whose presence can also be shown through the CPF
correlation.

From Eqs. (\ref{CPFSpin}) and (\ref{Gauss1}) $[f(t)\simeq \exp [-2(gt)^{2}],$
it follows the approximation%
\begin{equation}
C_{pf}(t,\tau )\simeq \frac{e^{-2g^{2}(t+\tau )^{2}}+e^{-2g^{2}(t-\tau )^{2}}%
}{2}-e^{-2g^{2}(t^{2}+\tau ^{2})}.  \label{CPFGaussianAprox}
\end{equation}%
For increasing number of spins $N,$ this expression provides a very well
fitting of Eq.~(\ref{CPFSpin}). In Fig. 1(b), we plot the correlation for
equal times, $C_{pf}(t,t).$ After a transient, it reaches a plateau regime, $%
C_{pf}(t,t)=1/2.$ This is the expected behavior when $N\rightarrow \infty .$
In fact, the correlation of the spin bath does not decay in time \cite{PRL}.
On the other hand, for finite $N$ recursive time behaviors are expected.
This property is clear seen in Fig. 2, where as in Fig. 1 we taken $N=50.$
Due to the natural recurrence time of the total unitary dynamics, the
temporal behavior is periodic in both measurement times (not shown).
Consistently, the localization of the central peak (around $gt=g\tau \simeq
5.5)$ goes to infinity for increasing $N.$

\section{Hamiltonian noise models}

The spin bath model [Eq. (\ref{SpinBath})] has not a natural Markovian
limit. In fact, the reservoir correlation does not decay in time. In solid
state environments extra degrees of freedom coupled to the spin variables
induce this feature. A simple way of representing this situation is to
approximate the spin bath and \textquotedblleft its
environment\textquotedblright\ by a classical colored noise \cite{anderson}.
Thus, it is considered the stochastic Hamiltonian evolution%
\begin{equation}
\frac{d}{dt}\rho _{t}^{st}=-i\xi (t)[\sigma _{\hat{z}},\rho _{t}^{st}],
\label{Hst}
\end{equation}%
where the system state\ $\rho _{t}$ follows by averaging $\rho _{t}^{st}$
over realizations of the real noise $\xi _{t},$ $\rho _{t}=\overline{\rho
_{t}^{st}}$ \cite{GaussianNoise}, which is denoted with the overbar symbol.

For simplicity, we only consider pure initial conditions. Hence, the problem
can be studied through a stochastic wave vector, $\rho _{t}^{st}=|\psi
_{t}\rangle \langle \psi _{t}|,$ whose evolution is%
\begin{equation}
\frac{d}{dt}|\psi _{t}\rangle =-i\xi (t)\sigma _{\hat{z}}|\psi _{t}\rangle .
\end{equation}%
Taking the initial state $|\psi _{0}\rangle =a|+\rangle +b|-\rangle ,$ which
is uncorrelated from the noise realizations, the stochastic wave vector reads%
\begin{equation}
|\psi _{t}\rangle =e^{-i\int_{0}^{t}dt^{\prime }\xi (t^{\prime })}a|+\rangle
+e^{+i\int_{0}^{t}dt^{\prime }\xi (t^{\prime })}b|-\rangle .
\label{PsiSolution}
\end{equation}%
This stochastic dynamics replaces the bipartite description given by Eq. (%
\ref{Psi(t)}).

Memory effects induced by the noise $\xi (t)$ can be studied through the
measurement scheme associated to the CPF correlation. Similarly to the spin
bath model, here the three successive measurements are chosen as projective
ones, being performed in $\hat{x}$-direction (in the qubit Bloch sphere).
The outcomes of each measurement are $x=\pm 1,$ $y=\pm 1,$ and $z=\pm 1,$
with measurement operators $\{\Omega _{x}\}=\{\Omega _{y}\}=\{\Omega
_{z}\}=\{\Pi _{\hat{x}=\pm 1}\},$ where $\Pi _{\hat{x}=\pm 1}=|\hat{x}_{\pm
}\rangle \langle \hat{x}_{\pm }|,$ with $|\hat{x}_{\pm }\rangle =(|+\rangle
\pm |-\rangle )/\sqrt{2}.$ The system initial condition is taken as $|\psi
_{0}\rangle =|+\rangle ,$ which in turn is statistically independent of the
noise.

\subsection{Conditional probabilities}

The following calculations are performed by taking into account a particular
noise realization. After the first $x$-measurement, the system state suffer
the transformation $|\psi _{0}\rangle \rightarrow |\psi _{0}^{x}\rangle =\Pi
_{\hat{x}=x}|\psi _{0}\rangle /\sqrt{\langle \psi _{0}|\Pi _{\hat{x}=x}|\psi
_{0}\rangle },$ delivering%
\begin{equation}
|\psi _{0}^{x}\rangle =\frac{|+\rangle +x|-\rangle }{\sqrt{2}},
\end{equation}%
where $x=\pm 1$ is the outcome of the measurement. The probability of both
options is $P_{st}(x)=\langle \psi _{0}|\Pi _{\hat{x}=x}|\psi _{0}\rangle
=1/2.$

In the next step, during a time interval $t$ the system evolves following
the dynamics (\ref{PsiSolution}),%
\begin{equation}
|\psi _{t}^{x}\rangle =\frac{1}{\sqrt{2}}\Big{[}e^{-i\int_{0}^{t}dt^{\prime
}\xi (t^{\prime })}|+\rangle +e^{+i\int_{0}^{t}dt^{\prime }\xi (t^{\prime
})}x|-\rangle \Big{]}.  \label{PhiStochX}
\end{equation}%
The probability for the second measurement outcomes $y=\pm 1$ follow from $%
P_{st}(y|x)=\langle \psi _{t}^{x}|\Pi _{\hat{x}=y}|\psi _{t}^{x}\rangle ,$
which then reads $P_{st}(y|x)=(1+yx\mathrm{Re}[\exp
[-2i\int_{0}^{t}dt^{\prime }\xi (t^{\prime })]])/2.$ Clearly, this object is
random and depends on each particular noise realization. The joint
probability distribution $P_{st}(y,x)=P_{st}(y|x)P_{st}(x),$ is given by%
\begin{equation}
P_{st}(y,x)=\frac{1}{4}(1+yx\mathrm{Re}[e^{-2i\int_{0}^{t}dt^{\prime }\xi
(t^{\prime })}]),  \label{NoisyP(x,y)}
\end{equation}%
which in turn implies, $P_{st}(y)=\sum_{x=\pm 1}P_{st}(y,x)=1/2.$ Thus, the
retrodicted probability $P_{st}(x|y)=P_{st}(y,x)/P_{st}(y)$ can be written as%
\begin{equation}
P_{st}(x|y)=\frac{1}{2}(1+yx\mathrm{Re}[e^{-2i\int_{0}^{t}dt^{\prime }\xi
(t^{\prime })}]).  \label{PRetroSt}
\end{equation}

After the second $y$-measurement, the wave vector collapses as $|\psi
_{t}^{x}\rangle \rightarrow |\psi _{t}^{yx}\rangle =\Pi _{\hat{x}=y}|\psi
_{t}^{x}\rangle /\sqrt{\langle \psi _{t}^{x}|\Pi _{\hat{x}=y}|\psi
_{t}^{x}\rangle },$ delivering%
\begin{equation}
|\psi _{t}^{yx}\rangle =\frac{|+\rangle +y|-\rangle }{\sqrt{2}}.
\end{equation}%
Notice that this state only depend on the last outcome $y,$ being
independent of the previous outcome $x.$ In addition it does not depend of
the measurement time $t,$ neither on the particular noise realization.

In the next step, the system evolves with the stochastic unitary evolution (%
\ref{PsiSolution}) during a time interval $\tau ,$ $|\psi _{t}^{yx}\rangle
\rightarrow |\psi _{t+\tau }^{yx}\rangle .$ Hence,%
\begin{equation}
|\psi _{t+\tau }^{yx}\rangle =\frac{1}{\sqrt{2}}\Big{[}e^{-i\int_{t}^{t+\tau
}dt^{\prime }\xi (t^{\prime })}|+\rangle +e^{+i\int_{t}^{t+\tau }dt^{\prime
}\xi (t^{\prime })}y|-\rangle \Big{]}.  \label{Intermedia}
\end{equation}%
The probability for the third $z$-measurement is $P_{st}(z|y,x)=\langle \psi
_{t+\tau }^{yx}|\Pi _{\hat{x}=z}|\psi _{t+\tau }^{yx}\rangle ,$%
\begin{equation}
P_{st}(z|y,x)=\frac{1}{2}(1+zy\mathrm{Re}[e^{-2i\int_{t}^{t+\tau }dt^{\prime
}\xi (t^{\prime })}]).  \label{PRetrozyx}
\end{equation}%
We notice that the conditional probabilities (\ref{PRetroSt}) and (\ref%
{PRetrozyx}) depend on each particular noise realization in the intervals $%
(0,t)$ and $(t,t+\tau ).$

\subsection{Dynamics between consecutive measurements}

The dynamics between consecutive measurements can be obtained by averaging
over noise realizations. After the first $x$-measurement and before the
second $y$-measurement, the system state follows as $\rho _{t}^{x}=\overline{%
|\psi _{t}^{x}\rangle \langle \psi _{t}^{x}|},$ where $|\psi _{t}^{x}\rangle 
$ is given by Eq.~(\ref{PhiStochX}). We get,%
\begin{equation}
\rho _{t}^{x}=\frac{1}{2}\left( 
\begin{array}{cc}
1 & xc_{t} \\ 
xc_{t}^{\ast } & 1%
\end{array}%
\right) ,  \label{RhoXNoise}
\end{equation}%
where the coherence behavior is given by%
\begin{equation}
c_{t}=\overline{c_{st}(t)}\equiv \overline{\exp \Big{[}-2i\int_{0}^{t}dt^{%
\prime }\xi (t^{\prime })\Big{]}}.  \label{c(t)Average}
\end{equation}%
Similarly to the quantum spin bath model, only the system coherences are
affected.

After the second $y$-measurement and before the third $z$-measurement, the
system state follows as $\rho _{t,\tau }^{yx}=\left. \overline{|\psi
_{t+\tau }^{yx}\rangle \langle \psi _{t+\tau }^{yx}|}\right\vert _{yx},$
where $|\psi _{t+\tau }^{yx}\rangle $ is given by Eq.~(\ref{Intermedia}). We
get,%
\begin{equation}
\rho _{t,\tau }^{yx}=\frac{1}{2}\left( 
\begin{array}{cc}
1 & yc_{t,\tau }^{yx} \\ 
yc_{t,\tau }^{yx\ast } & 1%
\end{array}%
\right) ,  \label{RhoYNoise}
\end{equation}%
where $c_{t,\tau }^{yx}$ is given by%
\begin{equation}
c_{t,\tau }^{yx}=\left. \overline{c_{st}(t,\tau )}\right\vert _{yx}\equiv
\left. \overline{\exp \Big{[}-2i\int_{t}^{t+\tau }dt^{\prime }\xi (t^{\prime
})\Big{]}}\right\vert _{yx}.  \label{CyxTauStoch}
\end{equation}

In contrast with Eq.~(\ref{c(t)Average}), in the previous expression the
classical average (denoted as $\left. \overline{\mathcal{F}[\xi ]}%
\right\vert _{yx},$ where $\mathcal{F}[\xi ]$ is a functional of the noise)
is restricted (conditioned) to the occurrence of the previous $x$ and $y$
measurement outcomes. The probability $P([\xi ]|y,x)$ of a noise realization
conditioned on these outcomes, from Bayes rule, is%
\begin{equation}
P([\xi ]|y,x)=\frac{P_{st}(y,x)P([\xi ])}{\overline{P_{st}(y,x)}}.
\label{BayesConditional}
\end{equation}%
Here, $P([\xi ])$ is the unconditional probability of a noise realization.
Furthermore, $P_{st}(y,x)$ is the joint probability of $y$ and $x$ outcomes
conditioned to a given noise realization. Thus, it is given by Eq. (\ref%
{NoisyP(x,y)}). For an arbitrary noise functional, the conditional average $%
\left. \overline{\mathcal{F}[\xi ]}\right\vert _{yx}$\ can then be written as%
\begin{equation}
\left. \overline{\mathcal{F}[\xi ]}\right\vert _{yx}=\overline{\mathcal{F}%
[\xi ]P([\xi ]|y,x)}=\frac{\overline{\mathcal{F}[\xi ]P_{st}(y,x)}}{%
\overline{P_{st}(y,x)}}.
\end{equation}%
Applying this result to Eq. (\ref{CyxTauStoch}), we get the final expression%
\begin{equation}
c_{t,\tau }^{yx}=\overline{\frac{c_{st}(t,\tau )(1+yx\mathrm{Re}[c_{st}(t)])%
}{1+yx\overline{\mathrm{Re}[c_{st}(t)]}}},  \label{CxyNoise}
\end{equation}%
which is defined in terms of unconditional classical ensemble averages.

We notice that $c_{t,\tau }^{yx}\neq c_{t}.$ \textit{This change of dynamics
follows from a measurement back action on the (average) environmental
influence.} In fact, this feature emerges from conditioning the classical
noise average to the occurrence of previous quantum measurement outcomes.
The different coherence behaviors indicates the non-Markovian property of
the system dynamics. In fact, Eqs. (\ref{c(t)Average}) and (\ref{CxyNoise})
are the analog of Eqs. (\ref{CoherTimeMicro}) and (\ref{CtTau}), which
correspond to the quantum spin environment.

\subsection{CPF correlation}

The conditional probabilities $P_{st}(x|y)$ [Eq. (\ref{PRetroSt})] and $%
P_{st}(z|y,x)$ [Eq.~(\ref{PRetrozyx})] relies on quantum measurement theory.
They were calculated taking into account a single noise realization. The
probability $P(z,x|y),$ which defines the CPF correlation (\ref{CPFExplicit}%
), describes the statistics for an ensemble of (conditional) measurement
results. Given its conditional character, it can be written as%
\begin{equation}
P(z,x|y)=\overline{P_{st}(z,x}|y)=\overline{P_{st}(z|y,x)P_{st}(x|y)}|_{y},
\label{CPFProbConditional}
\end{equation}%
where the (classical noise) average is conditioned to the occurrence of a
particular $y$-outcome. This average can be performed with the conditional
probability $P([\xi ]|y)=P_{st}(y)P([\xi ])/\overline{P_{st}(y)},$ where $%
P_{st}(y)$ is the probability of obtaining a $y$-outcome for a given noise
realization. Nevertheless, for the chosen initial conditions $%
P_{st}(y)=\sum_{x=\pm 1}P_{st}(y,x)=1/2$ [see Eq. (\ref{NoisyP(x,y)})].
Thus, $P([\xi ]|y)=P([\xi ]),$ which implies that, for the chosen initial
conditions, the noise average in Eq. (\ref{CPFProbConditional}) can be taken
as an unconditional one, $P(z,x|y)=\overline{P_{st}(z|y,x)P_{st}(x|y)}.$
Using this result, from Eqs. (\ref{PRetroSt}) and (\ref{PRetrozyx}) it is
possible to obtain%
\begin{equation}
P(z,x|y)=\frac{1}{4}\Big{[}1+xyf(t)+zyf^{\prime }(\tau )+zxf(t,\tau )\Big{]}.
\label{CPFNoiseAveraged}
\end{equation}%
The auxiliary functions are%
\begin{equation}
f(t)=\overline{\mathrm{Re}[c_{st}(t)]}=\overline{\mathrm{Re}\left[
e^{-2i\int_{0}^{t}dt^{\prime }\xi (t^{\prime })}\right] }.  \label{h(t)}
\end{equation}%
Thus, $f(t)=\mathrm{Re}[c_{t}]$ [Eq. (\ref{c(t)Average})]. Furthermore,%
\begin{equation}
f^{\prime }(\tau )=\overline{\mathrm{Re}[c_{st}(t,\tau )]}=\overline{\mathrm{%
Re}\left[ e^{-2i\int_{t}^{t+\tau }dt^{\prime }\xi (t^{\prime })}\right] }.
\label{ftauprima}
\end{equation}%
For \textit{stationary noises} \cite{vanKampen} this function does not
depend on the time $t.$ In fact, stationarity implies $f^{\prime }(\tau
)=f(\tau ).$ Finally,%
\begin{equation}
f(t,\tau )=\overline{\mathrm{Re}[c_{st}(t)]\mathrm{Re}[c_{st}(t,\tau )]},
\end{equation}%
which explicitly reads%
\begin{equation}
f(t,\tau )=\overline{\mathrm{Re}\left[ e^{-2i\int_{0}^{t}dt^{\prime }\xi
(t^{\prime })}\right] \mathrm{Re}\left[ e^{-2i\int_{t}^{t+\tau }dt^{\prime
}\xi (t^{\prime })}\right] }.  \label{h(t,tau)}
\end{equation}

From Eq. (\ref{CPFNoiseAveraged}), using that $O_{z}=z,$ $O_{x}=x,$ the
conditional averages read $\overline{\langle O_{z}\rangle _{y}}=yf(\tau ),$ $%
\overline{\langle O_{x}\rangle _{y}}=yf(t),$ and $\overline{\langle
O_{z}O_{x}\rangle _{y}}=f(t,\tau ).$ The CPF correlation, $C_{pf}(t,\tau )=%
\overline{\langle O_{z}O_{x}\rangle _{y}}-\overline{\langle O_{z}\rangle _{y}%
}\ \overline{\langle O_{x}\rangle _{y}},$ becomes%
\begin{equation}
C_{pf}(t,\tau )=f(t,\tau )-f(t)f(\tau ).  \label{CPFexactNoise}
\end{equation}%
These expressions recover the exact results presented in Ref. \cite{PRL}.

Taking into account the expressions (\ref{h(t)}) and (\ref{h(t,tau)}), we
realize that $C_{pf}(t,\tau )$ corresponds to the centered correlation of
the real part of the phase terms $\exp [-2i\int_{0}^{t}dt^{\prime }\xi
(t^{\prime })]$ and $\exp [-2i\int_{t}^{t+\tau }dt^{\prime }\xi (t^{\prime
})],$ which in turn correspond to the (stochastic) coherence decay behaviors
in the intervals $(0,t)$ and $(t,t+\tau ),$ respectively.

The exact result (\ref{CPFexactNoise}) can be evaluated for arbitrary
(stationary) noises. The (real) functions $f(t)$ and $f(t,\tau )$ also
determine the system dynamics between consecutive measurements, Eqs. (\ref%
{c(t)Average}) and (\ref{CxyNoise}). In fact, for a stationary noise with a
vanishing mean value, $\overline{\xi (t)}=0,$ it follows%
\begin{equation}
c_{t}=f(t),\ \ \ \ \ \ \ \ \ \ \ \ c_{t,\tau }^{yx}=\frac{f(\tau
)+yxf(t,\tau )}{1+yxf(t)}.  \label{RealCoherenciasNoise}
\end{equation}

\subsection{Gaussian Noise}

Gaussian fluctuations arise naturally in different physical situations such
as for example\ in solid state environments \cite{anderson,abragam}. For
this statistics, the calculation of the functions $f(t)$ and $f(t,\tau )$
[Eqs. (\ref{h(t)}) and (\ref{h(t,tau)})] can be performed in different
alternative ways. Here, they are determine through the characteristic noise
functional \cite{vanKampen},%
\begin{equation}
G[k]=\overline{\exp \left[ i\int_{0}^{\infty }k(t^{\prime })\xi (t^{\prime
})dt^{\prime }\right] },
\end{equation}%
which depends on an arbitrary test function $k(t).$ For a Gaussian noise
with null average, $\overline{\xi (t)}=0,$ it reads \cite{vanKampen}%
\begin{equation}
G[k]=\exp \Big{[}-\frac{1}{2}\int_{0}^{\infty }dt_{2}\int_{0}^{\infty
}dt_{1}k(t_{2})k(t_{1})\chi (t_{2},t_{1})\Big{]},  \label{G[k]Gauss}
\end{equation}%
where $\chi (t_{2},t_{1})\equiv \overline{\xi (t_{2})\xi (t_{1})}=\chi
(|t_{2}-t_{1}|)$ is the noise correlation. The last equality is valid for
stationary noises.

After giving an explicit noise correlation, the averages (\ref{h(t)}) and (%
\ref{h(t,tau)}) follows by writing $\mathrm{Re}[a]=(a+a^{\ast })/2,$ and by
taking an adequate set of test functions. For example, $\overline{\exp
i\int_{0}^{t}\xi (t^{\prime })dt^{\prime }}$ follows from $G[k]$ with $%
k(t^{\prime })=\theta (t-t^{\prime }),$ where $\theta (x)$ is the step
function, and performing the corresponding time integrals. Similarly, the
calculus of $\overline{\exp i[\int_{t}^{t+\tau }\xi (t^{\prime })dt^{\prime
}\pm \int_{0}^{t}\xi (t^{\prime })dt^{\prime }]}$ is obtained with $%
k(t^{\prime })=\theta (t+\tau -t^{\prime })\theta (t^{\prime }-t)\pm \theta
(t-t^{\prime }).$

\subsubsection{White noise}

For a white noise, $\chi (t_{2},t_{1})=\gamma _{w}\delta (t_{2}-t_{1}),$ it
follows%
\begin{equation}
f(t)=\exp [-2\gamma _{w}t],\ \ \ \ \ \ \ \ \ f(t,\tau )=f(t)f(\tau ).
\label{FunctionsMarkov}
\end{equation}%
Hence, a Markovian limit is achieved,%
\begin{equation}
c_{\tau }=c_{t,\tau }^{yx}=f(\tau ),\ \ \ \ \ \ C_{pf}(t,\tau )=0.
\label{CPFMarkov}
\end{equation}%
In fact, here the system dynamics between consecutive measurement events do
not depend on the measurement outcomes and are the same. Consistently, the
CPF correlation vanishes. Furthermore, both intermediate dynamics [Eqs. (\ref%
{RhoXNoise}) and (\ref{RhoYNoise})] obey a (dephasing) Lindblad evolution,%
\begin{equation}
\frac{d\rho _{t}}{dt}=\frac{1}{2}\gamma (t)(\sigma _{\hat{z}}\rho _{t}\sigma
_{\hat{z}}-\rho _{t}),  \label{LindbladTime}
\end{equation}%
where the time dependent rate is determined by the coherence behavior%
\begin{equation}
\gamma (t)=-\frac{1}{\langle +|\rho _{t}|-\rangle }\frac{d}{dt}\langle
+|\rho _{t}|-\rangle .  \label{rate}
\end{equation}%
Thus, in both cases $\gamma (t)=2\gamma _{w}.$

\subsubsection{Infinite correlation-time}

This case corresponds to a noise correlation that does not decay in time, $%
\chi (t_{2},t_{1})=g^{2}.$ We obtain%
\begin{equation}
f(t)=\exp [-2(gt)^{2}],\ \ \ \ \ f(t,\tau )=f(t+\tau )+f(t-\tau ).
\label{FunctionsInfiniteCorre}
\end{equation}%
This decay behavior recovers the (asymptotic in $N)$ dynamics induced by the
spin bath model when developing pure Gaussian decay behaviors. Thus, the
exact expression for the coherences $c_{\tau }$ and $c_{t,\tau }^{yx}$ [Eq.~(%
\ref{RealCoherenciasNoise})] are given by Eqs. (\ref{Gauss1}) and (\ref%
{Gauss2}) respectively. The CPF correlation then correspond to Eq. (\ref%
{CPFGaussianAprox}). Between the first two measurements\ [Eq. (\ref%
{RhoXNoise})], the system evolution is given by Eq.~(\ref{LindbladTime})
with $\gamma (t)=4g^{2}t>0.$ A more complex expression (which depends on the
product $yx$ of measurement outcomes) describes the rate for the second
evolution [Eq.~(\ref{RhoYNoise})].

\subsubsection{Exponential correlation}

For an exponential correlation%
\begin{equation}
\chi (t_{2},t_{1})=g^{2}\exp [-|t_{2}-t_{1}|/\tau _{c}],
\label{ExponentialCorre}
\end{equation}%
where the parameter $\tau _{c}$\ gives the characteristic correlation-time
of the noise fluctuations, it follows%
\begin{equation}
f(t)=\exp \Big{\{}-4(\tau _{c}g)^{2}[t/\tau _{c}-(1-e^{-t/\tau _{c}})]%
\Big{\}}.  \label{h(t)ExpoCorr}
\end{equation}%
The function (\ref{h(t,tau)}) reads%
\begin{equation}
f(t,\tau )=f(t)f(\tau )\cosh [\varphi (t,\tau )],  \label{f2TiemposNoise}
\end{equation}%
where the auxiliary function is%
\begin{equation}
\varphi (t,\tau )\equiv 4(\tau _{c}g)^{2}(1-e^{-t/\tau _{c}})(1-e^{-\tau
/\tau _{c}}).
\end{equation}

In the limit $\tau _{c}\rightarrow \infty ,$ these expressions consistently
give the infinite correlation limit (\ref{FunctionsInfiniteCorre}). On the
other hand, taking $\gamma _{w}/2=g^{2}\tau _{c}$ as a constant parameter,
in the limit $\tau _{c}\rightarrow 0,$ the Markovian regime\ (\ref%
{FunctionsMarkov}) is recovered.

In Fig. 3 we plot the coherence decay $c_{\tau }$ and $c_{t,\tau }^{yx}$
[Eq.~(\ref{RealCoherenciasNoise})] between consecutive measurement events.
For larger correlations times $(g\tau _{c}=100),$ the behavior is similar to
that of the quantum spin bath (see Fig. 1). On the other hand, for smaller
correlation times $(g\tau _{c}=5),$ the measurement back action on the
coherence behavior is diminished. In fact, the difference between both
dynamics disappear in a white noise limit. The dynamics between the first
two measurement is given by the Lindblad evolution (\ref{LindbladTime}) with 
$\gamma (t)=4g^{2}\tau _{c}(1-e^{-t/\tau _{c}})>0,$ while a more complex
expression describes the rate for the second evolution [Eq. (\ref{RhoYNoise}%
)].

In Fig. 4, for the same correlation time $(g\tau _{c}=5),$ it is plotted the
CPF correlation [Eq.~(\ref{CPFexactNoise})]. Its maximal amplitude (central
peak) diminishes when $\tau _{c}$ diminishes (compare with Fig. 2, which can
be read as the $\tau _{c}\rightarrow \infty $ limit). Furthermore, as the
Markovian limit is being approached, $C_{pf}(t,\tau )$ is not null only at
short times.

In order to visualize the transition between Markovian and non-Markovian
regimes, in Fig. 5(a) we plot the coherence behavior $c_{\tau }$ [Eq.~(\ref%
{RealCoherenciasNoise})]. Different values of the correlation $\tau _{c}$
are chosen while the parameter $\gamma _{w}/2=g^{2}\tau _{c}$ remains
constant. A transition between exponential and Gaussian behaviors is clearly
seen when increasing the correlation time $\tau _{c}.$ In Fig. 5(b) we also
plot the CPF correlation for a set of different correlation times. For
increasing $\tau _{c}$ the limit (\ref{CPFGaussianAprox}) is recovered,
while for decreasing $\tau _{c}$ the CPF correlation approaches the
Markovian limit (\ref{CPFMarkov}). 
\begin{figure}[tbp]
\includegraphics[bb=45 871 750 1132,angle=0,width=8.8cm]{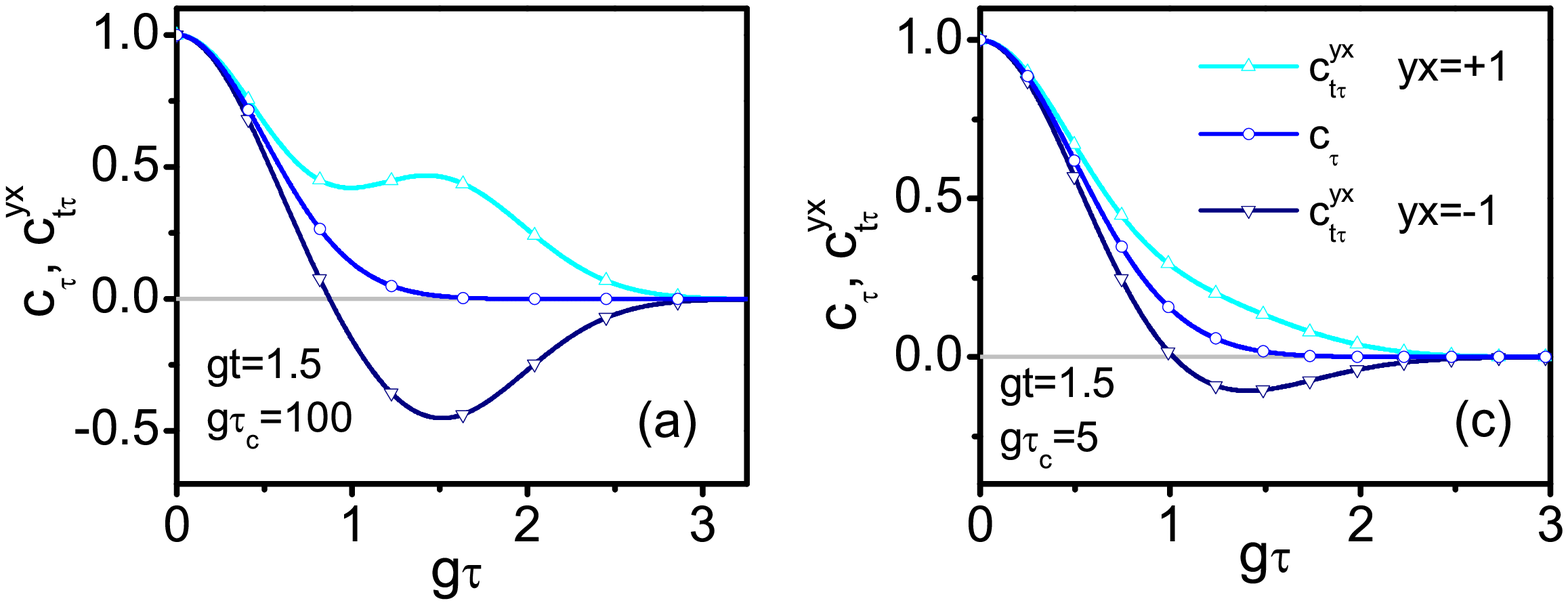}
\caption{Coherence decay behaviors $c_{\protect\tau }$ and $c_{t,\protect%
\tau }^{yx}$ between consecutive measurements [Eq. (\protect\ref%
{RealCoherenciasNoise})] for the Gaussian stochastic Hamiltonian model with
exponential correlation (\protect\ref{ExponentialCorre}) for different
correlation times $\protect\tau _{c},$ (a) $g\protect\tau _{c}=100,$ (b) $g%
\protect\tau _{c}=5.$ In both cases $gt=1.5.$}
\end{figure}

\begin{figure}[t]
\includegraphics[bb=0 0 340 252,angle=0,width=7cm]{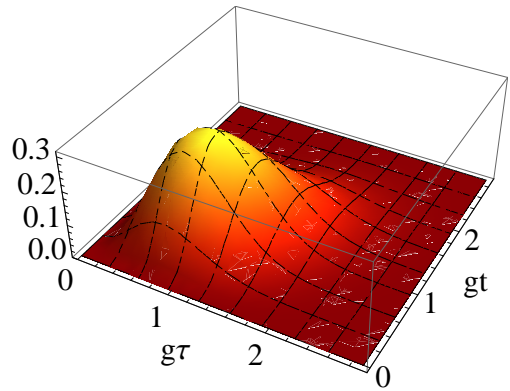}
\caption{CPF correlation [Eq. (\protect\ref{CPFexactNoise}) jointly with
Eqs.~(\protect\ref{h(t)ExpoCorr}) and (\protect\ref{f2TiemposNoise})] for
the Gaussian stochastic Hamiltonian model with exponential correlation (%
\protect\ref{ExponentialCorre}) with parameters $g\protect\tau _{c}=5.$}
\end{figure}

\begin{figure}[b]
\includegraphics[bb=51 866 750 1132,angle=0,width=8.8cm]{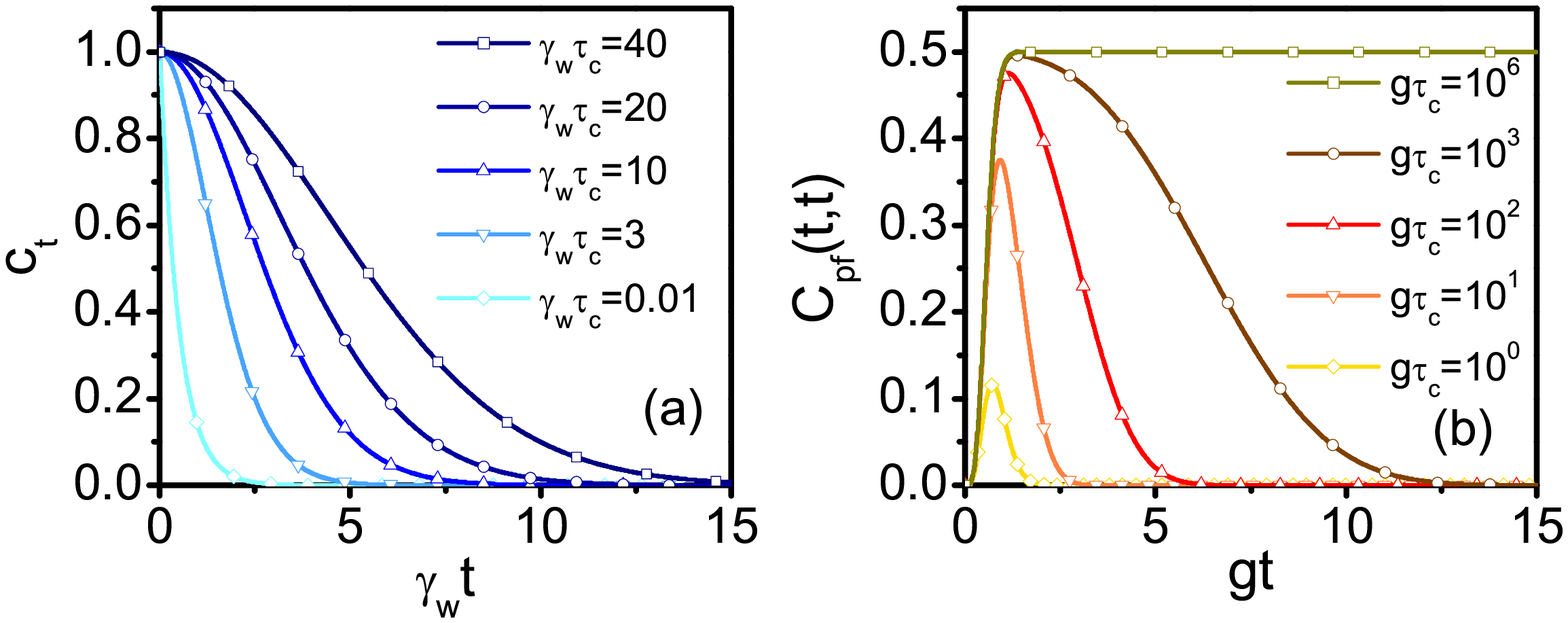}
\caption{(a) System coherence $c_{\protect\tau }$ [Eqs. (\protect\ref%
{RealCoherenciasNoise}) and (\protect\ref{h(t)ExpoCorr})] for the Gaussian
stochastic Hamiltonian model with exponential correlation (\protect\ref%
{ExponentialCorre}) for different correlation times $\protect\tau _{c},$
where $\protect\gamma _{w}/2=g^{2}\protect\tau _{c}.$ (b) CPF correlation (%
\protect\ref{CPFexactNoise}) for equal times, $C_{pf}(\protect\tau ,\protect%
\tau ),$ for different correlation times $\protect\tau _{c}.$}
\end{figure}

\section{Non-Markovian Dephasing Lindblad evolutions}

Lindblad dynamics [like Eq. (\ref{LindbladTime})] with positive rates are
associated to a Markovian regime \cite{BreuerReview,plenioReview}. In the
present scheme quantum Markovianity does not rely on Lindblad theory. It is
defined by a vanishing CPF\ correlation. If both evolutions between
consecutive measurement events are defined by the same Lindblad equation the
CPF correlation vanishes. In the previous Hamiltonian model this situation
arises when the noise is a delta correlated one.

In this section we show that even when the evolution between the first and
second measurement events is given by a time-independent Lindblad equation,
the posterior evolution (between the second and third measurements) may
change, implying a non-vanishing CPF correlation. Thus, the original
Lindblad equation cannot be associated to a Markovian dynamics. This unusual
non-Markovian effect emerges when the underlying parameters of the studied
models become Lorentzian random variables \cite{vanKampen,caceres}.

\subsection{Spin environment with random coupling}

In solid state environments the couplings $\{g_{k}\}$ in the spin bath model
(\ref{SpinBath}) may become random variables \cite{Paz}. This feature may
represents, for example, distance-dependent system-bath interactions
modulated by the random location of each spin of the environment \cite%
{abragam}. Independently of its physical origin, the description of a random
coupling model follows from the results of Sec.~II after averaging over the
distribution of the set $\{g_{k}\}:$%
\begin{equation}
\mathbf{c}_{t}\equiv \overline{c_{t}},\ \ \ \ \ \ \mathbf{c}_{t,\tau
}^{yx}\equiv \left. \overline{c_{t,\tau }^{yx}}\right\vert _{yx},\ \ \ 
\mathbf{C}_{pf}(t,\tau )\equiv \left. \overline{C_{pf}(t,\tau )}\right\vert
_{y}.  \label{DefRandomAverages}
\end{equation}%
In these expressions bold letters denote averaged quantities. The (random)
objects are defined by Eqs. (\ref{CoherTimeMicro}), (\ref{CtTau}), and (\ref%
{CPFSpin}), evaluated in a particular realization of the set $\{g_{k}\}.$
The overbar denotes average over their probability distributions.

The average that gives $\mathbf{c}_{t}$ is an unconditional one.
Nevertheless, for$\ \mathbf{c}_{t,\tau }^{yx}$\ the classical average is
conditioned to the occurrence of $y$ and $x$ outcomes. Similarly to Eq.~(\ref%
{BayesConditional}), from Bayes rule this conditional average is defined by
the distribution%
\begin{equation}
P(\{g_{k}\}|y,x)=\frac{P(y,x)P(\{g_{k}\})}{\overline{P(y,x)}},
\end{equation}%
where $P(\{g_{k}\})$ is the (unconditional) probability distribution of the
coupling constants, while $P(y,x)$ is given by Eq. (\ref{ConjuntaYXSpin})
evaluated in a particular realization of the set $\{g_{k}\}.$ Thus, from Eq.
(\ref{CtTau}), the coherence behavior between the first two ($x$ and $y$)
measurements is%
\begin{equation}
\mathbf{c}_{t,\tau }^{yx}=\overline{\frac{c_{\tau }+yx(c_{t+\tau }+c_{t-\tau
}^{\ast })/2}{1+yx\overline{(c_{t}+c_{t}^{\ast })/2}}},  \label{CxyLorentz}
\end{equation}%
which is written in terms of unconditional averages.$\ $

The correlation $\mathbf{C}_{pf}(t,\tau )$ [Eq. (\ref{DefRandomAverages})]
is defined by a classical average conditioned to the occurrence of a
particular $y$-outcome. Nevertheless, due to the chosen initial conditions,
similarly to the average in Eq. (\ref{CPFProbConditional}), it can be taken
as an unconditional one. Thus, $\mathbf{C}_{pf}(t,\tau )=\overline{%
C_{pf}(t,\tau )}.$

\subsubsection*{Lorentz probability distribution}

The\textit{\ coupling }$\{g_{k}\}$\textit{\ are taken as independent
identical random variables}, with the scaling%
\begin{equation}
g_{k}=\frac{1}{N}\tilde{g}.  \label{LorentzScaling}
\end{equation}%
The probability density of\ the random variable $\tilde{g}$ is a Lorentzian
one,%
\begin{equation}
P(\tilde{g})=\frac{\gamma /2}{\pi \lbrack (\tilde{g}-\omega /2)^{2}+(\gamma
/2)^{2}]},  \label{Lorentz}
\end{equation}%
where $\gamma $ and $\omega $ are free parameters. Denoting with an overbar
the average over $\tilde{g},$ it follows the relation \cite{caceres}%
\begin{equation}
\overline{\exp (+2i\tilde{g}t)}=\int_{-\infty }^{+\infty }d\tilde{g}P(\tilde{%
g})e^{+2i\tilde{g}t}=\exp (i\omega t)\exp (-\gamma |t|).
\label{PromedioExpLorentz}
\end{equation}%
Thus, random phases with a Lorentzian distribution leads to exponential
decay behaviors.

Assuming that all spin of the reservoir begin in the same state, $\alpha
_{k}=\alpha ,$ $\beta _{k}=\beta ,$ with $|\alpha |^{2}+|\beta |^{2}=1,$
from Eq. (\ref{CoherTimeMicro}) the average coherence behavior $\mathbf{c}%
_{t}$ is given by%
\begin{equation}
\mathbf{c}_{t}=\overline{c_{t}}=e^{-\gamma |t|}(|\alpha |^{2}e^{+i\omega
t/N}+|\beta |^{2}e^{-i\omega t/N})^{N}.  \label{CLorentz}
\end{equation}%
Hence, and exponential decay behavior is valid for arbitrary $N.$
Furthermore, for $N\gg 1,$ it can be approximated as $\mathbf{c}_{t}\simeq
e^{-\gamma |t|}\exp [i(|\alpha |^{2}-|\beta |^{2})\omega t].$ If $|\alpha
|^{2}=|\beta |^{2}=1/2,$ or alternatively taking\ $\omega =0,$ the induced
complex phase vanishes. Thus, from Eq.~(\ref{CLorentz}) it follows the pure
exponential decay behavior%
\begin{equation}
\mathbf{c}_{t}=\exp [-\gamma |t|].  \label{CExponential}
\end{equation}%
Similarly, from Eq. (\ref{CxyLorentz}) it follows the exact result%
\begin{equation}
\mathbf{c}_{t,\tau }^{yx}=\frac{e^{-\gamma |\tau |}+yx(e^{-\gamma |t+\tau
|}+e^{-\gamma |t-\tau |})}{1+yxe^{-\gamma |t|}}.  \label{CYXRandomExplicit}
\end{equation}%
%
%
%
%
%
%
%
%
%
%
\begin{figure}[tbp]
\includegraphics[bb=45 870 734 1140,angle=0,width=8.5cm]{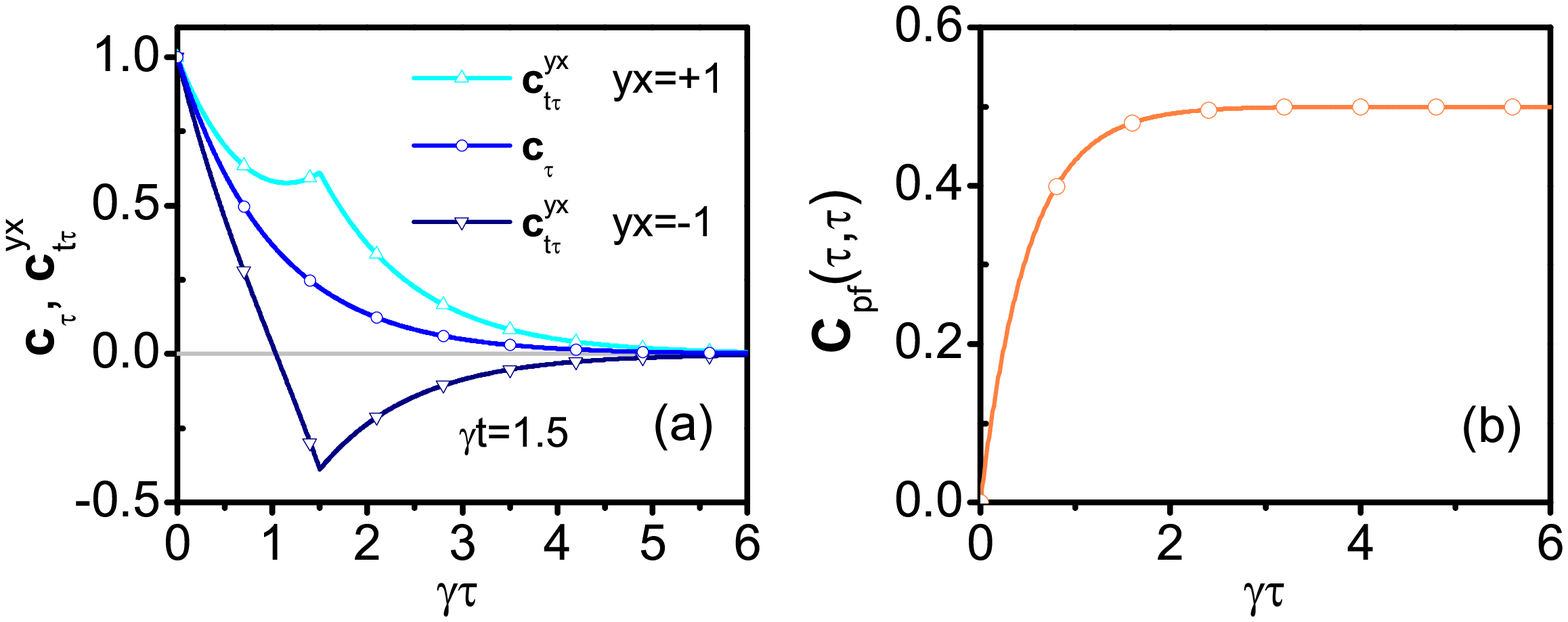}
\caption{(a) System coherences $\mathbf{c}_{t}$ [Eq. (\protect\ref%
{CExponential})] and $\mathbf{c}_{t,\protect\tau }^{yx}$\ [Eq. (\protect\ref%
{CYXRandomExplicit})] for the spin bath model with Lorentzian distributed
coupling. (b) CPF correlation (\protect\ref{CPFExponential}) for equal
times, $\mathbf{C}_{pf}(\protect\tau ,\protect\tau ).$ In both figures, the
random coupling are given by the scaling Eq. (\protect\ref{LorentzScaling}),
while the average coupling is null, $\protect\omega =0$ [Eq. (\protect\ref%
{Lorentz})], and $N=50.$ The system and bath initial conditions are the same
than in Fig.~1.}
\end{figure}

The exponential behavior (\ref{CExponential}) implies that between the first
two measurements the system dynamics is given by a dephasing Lindblad
equation with a time-independent rate [Eq. (\ref{LindbladTime}) with $\gamma
(t)=\gamma $]. Nevertheless, the second $y$-measurement induces a posterior
change of system behavior [see Eqs. (\ref{Psi0X}) and (\ref{Uncorrelato})].
The change $\mathbf{c}_{t}\rightarrow \mathbf{c}_{t,\tau }^{yx},$ in spite
of the former pure exponential behavior, indicates that the dynamics is
non-Markovian. Consequently,\textit{\ a Linblad dynamics does not guarantee
quantum Markovianity}. In Fig. 6(a) we show the behavior of both $\mathbf{c}%
_{t}$ and $\mathbf{c}_{t,\tau }^{yx},$ which is given by the previous two
expressions. $\mathbf{c}_{t,\tau }^{yx}$ develops a non-differentiable
time-behavior which is induced by the Lorentzian coupling statistics.

The non-Markovian property of the system dynamics can also be shown through
the CPF correlation. From Eqs. (\ref{CPFSpin}) and (\ref{DefRandomAverages})
[with $\mathbf{C}_{pf}(t,\tau )=\overline{C_{pf}(t,\tau )}]$
straightforwardly it follows the exact expression%
\begin{equation}
\mathbf{C}_{pf}(t,\tau )=\frac{e^{-\gamma |t+\tau |}+e^{-\gamma |t-\tau |}}{2%
}-e^{-\gamma (|t|+|\tau |)},  \label{CPFExponential}
\end{equation}%
which certainly is not null.

In Fig. 6(b) we plot $\mathbf{C}_{pf}(t,\tau )$ for equal times intervals
while in Fig. 7 we plot its dependence on both times. In contrast to Fig. 2,
due to the randomness of the coupling coefficients, the time behavior is not
periodic in time. Furthermore, the asymptotic behavior $\lim_{t\rightarrow
\infty }C_{pf}(t,t)=1/2$ again is related to an infinite environment
correlation-time.

The dynamics characterized previously demonstrates that a Lindblad equation
may arises even when the Born-Markov approximation does not applies. Notice
that the non-Markovian character of the evolution can only be detected
through extra information that is not encoded in the density matrix dynamics
corresponding to the time interval $(0,t).$ 
\begin{figure}[tbp]
\includegraphics[bb=0 0 340 260,angle=0,height=3.5cm]{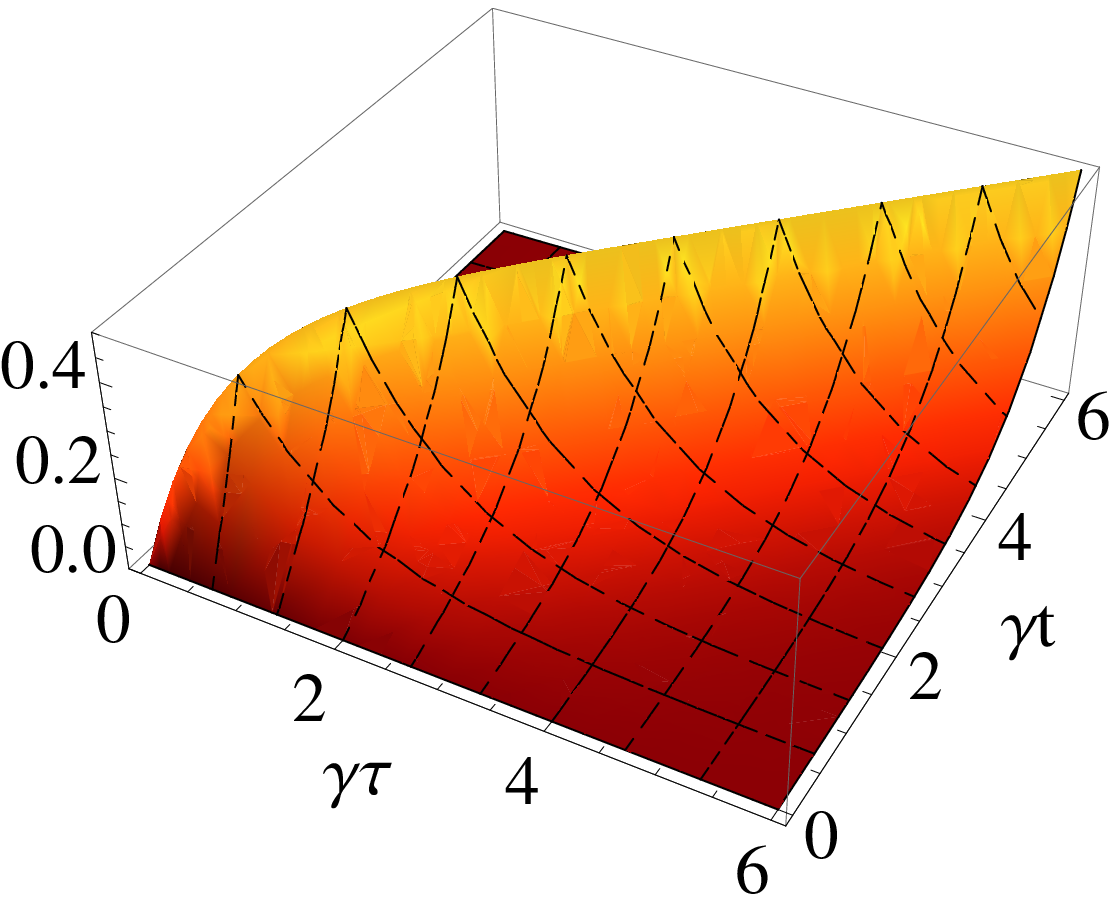} %
\includegraphics[bb=0 0 330 320,angle=0,height=3.5cm]{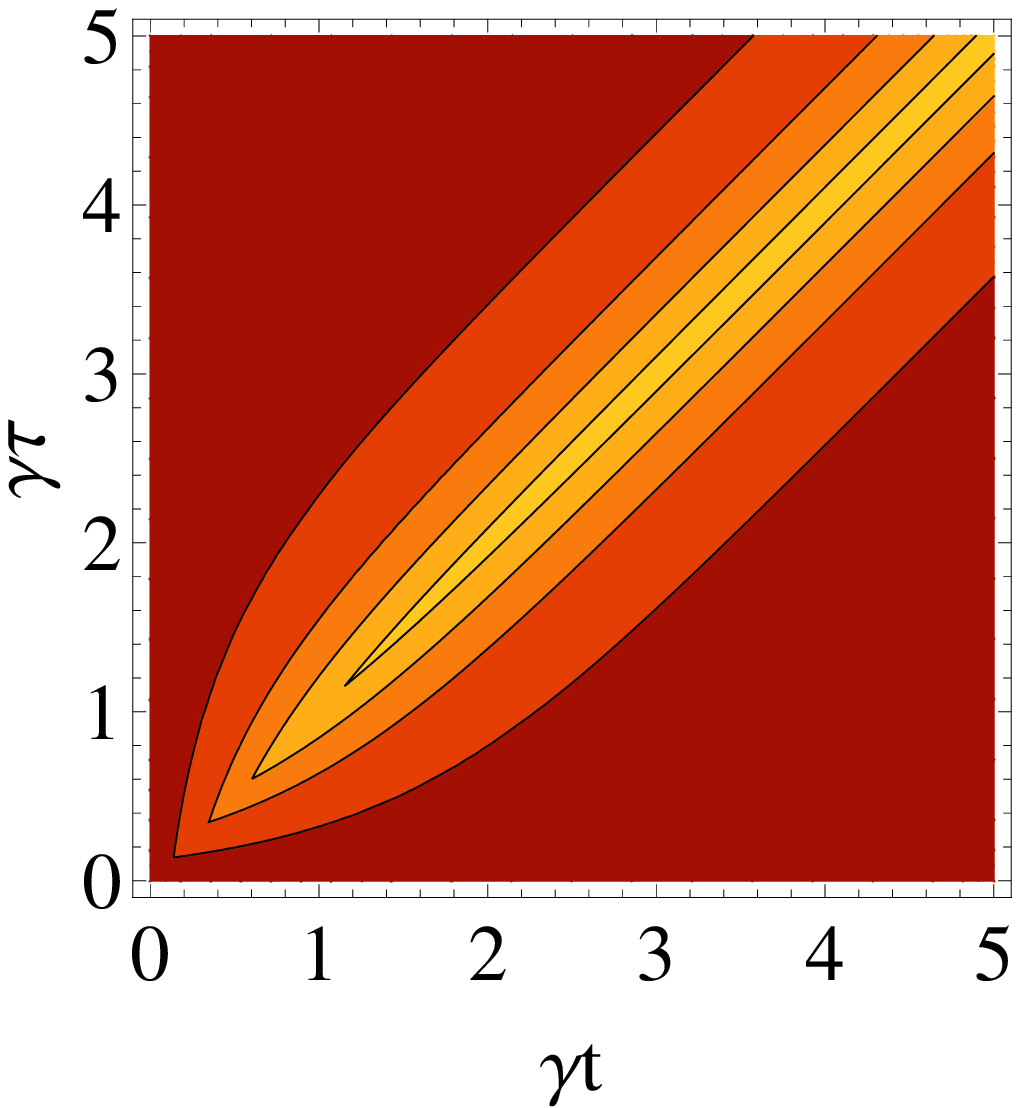}
\caption{CPF correlation $\mathbf{C}_{pf}(t,\protect\tau )$ [Eq. (\protect
\ref{CPFExponential})] for the spin bath model with Lorentzian coupling
constants. The parameters are the same than in the previous figure.}
\end{figure}

\subsection{Random frequency models}

Instead of a time-dependent stochastic noise one can consider a random
frequency model, that is, Eq. (\ref{Hst}) under the replacement $\xi
(t)\rightarrow \tilde{g},$%
\begin{equation}
\frac{d}{dt}\rho _{t}^{st}=-i\tilde{g}[\sigma _{\hat{z}},\rho _{t}^{st}],
\label{LorentzHamilton}
\end{equation}%
where $\tilde{g}$ is a (time-independent) random variable with probability
density $P(\tilde{g}).$ The infinite correlation limit of the Gaussian noise
[Eq. (\ref{G[k]Gauss})] can be read in this way, where $P(\tilde{g})$ is a
Gaussian distribution. On the other hand, we notice that the evolution (\ref%
{LorentzHamilton}) corresponds to a particular case of a (quantum-classical)
generalized Lindblad equation \cite{LindbladRate}.

All calculations performed in Sec. IV applies to the present model after
replacing $\xi (t)\rightarrow \tilde{g}.$ The functions $f(t)$ and $%
f^{\prime }(\tau )$ [Eqs. (\ref{h(t)}) and (\ref{ftauprima})] become 
\begin{equation}
f(t)=\frac{\overline{e^{+2i\tilde{g}t}+e^{-2i\tilde{g}t}}}{2},\ \ \ \
f^{\prime }(\tau )=\frac{\overline{e^{+2i\tilde{g}\tau }+e^{-2i\tilde{g}\tau
}}}{2},
\end{equation}%
where the overbar here denotes average with the distribution $P(\tilde{g}).$
Eq. (\ref{h(t,tau)}) becomes%
\begin{equation}
f(t,\tau )=\frac{\overline{(e^{+2i\tilde{g}t}+e^{-2i\tilde{g}t})(e^{+2i%
\tilde{g}\tau }+e^{-2i\tilde{g}\tau })}}{4}.
\end{equation}

\subsubsection*{Lorentzian random frequencies}

Similarly to the spin bath model, here we chose a Lorentzian distribution (%
\ref{Lorentz}) for $\tilde{g}.$ Taking $\omega =0,$ from Eq. (\ref%
{PromedioExpLorentz}) it follows%
\begin{equation}
f(t)=\exp [-\gamma |t|],\ \ \ \ f^{\prime }(\tau )=\exp [-\gamma |\tau |],
\end{equation}%
and similarly%
\begin{equation}
f(t,\tau )=\frac{\exp [-\gamma |t+\tau |]+\exp [-\gamma |t-\tau |]}{2}.
\end{equation}%
With these expressions at hand it is simple to realize that $c_{t}$ and $%
c_{t,\tau }^{yx}$ [Eq.(\ref{RealCoherenciasNoise})] are given by Eqs. (\ref%
{CExponential}) and (\ref{CYXRandomExplicit}) respectively. Furthermore, the
CPF correlation $C_{pf}(t,\tau )$ [Eq.~(\ref{CPFexactNoise})] is given by
Eq. (\ref{CPFExponential}). Therefore, the random frequency model leads to
the same results and expressions than\ the spin bath model with Lorentzian
random coefficients. This simplified model [Eq. (\ref{LorentzHamilton})]
also demonstrate that a Lindblad equation may relies on strong
system-environment correlations.

\section{Conclusions}

Similarly to classical systems, quantum non-Markovian effects can be studied
through a CPF correlation. Its definition relies on three quantum
measurements performed successively over the system of interest. We
characterized the CPF correlation for a qubit system whose non-Markovian
dynamics is induced by different dephasing mechanisms. Over the basis of
standard quantum measurement theory, exact expressions were found for a
quantum spin environment as well as for stochastic Hamiltonians models.

The present analysis allowed us to relate the presence of memory effects,
indicated by a nonvanishing CPF correlation, with a measurement back action
that change the system dynamics between consecutive measurement events. In
fact, in a Markovian limit, defined by a vanishing CPF correlation, this
dynamical change is absent. For the Hamiltonian noise model Markovianity
emerges in a white noise limit.

Taking the underlying parameters of the models as random variables with a
Lorentzian probability density, the former system evolution between the
first two measurements is given a dephasing Lindblad equation with a
time-independent rate. In spite of this feature, the posterior system
evolution, between the second and third measurements, is different from the
former one. This unexpected (non-Markovian) property demonstrates that
Lindblad equations may emerge even when the system and the environment are
highly correlated. Quantum non-Markovian measures based solely on the system
density matrix evolution are unable to detect these non-Markovian features.

\section*{Acknowledgments}

This work was supported by Consejo Nacional de Investigaciones Cient\'{\i}%
ficas y T\'{e}cnicas (CONICET), Argentina.

\appendix

\end{document}